\documentclass[pra,floatfix,twocolumn,superscriptaddress]{revtex4-1}
\usepackage[final]{graphicx}
\usepackage{times,bbm,amsmath,amssymb}
\usepackage{epsfig,color}
\usepackage{hyperref}
\usepackage{float,siunitx}
\usepackage[caption = false]{subfig}
\usepackage[greek,english]{babel}
\usepackage{thumbpdf,enumerate}
\usepackage{booktabs}
\usepackage{sidecap}
\usepackage[scaled=.8]{couriers}
\usepackage{pstricks}
\usepackage{multirow}
\usepackage{placeins}
\usepackage{pst-grad}
\usepackage{epigraph}
\usepackage{longtable}
\usepackage{booktabs}
\usepackage{gensymb}

\begin{document}

\preprint{APS/123-QED}

\title{Experimental method for measuring classical concurrence of generic beam shapes}

\author{E. Roccia}
\affiliation{Dipartimento di Scienze, Universit\`a degli Studi Roma Tre, Via della Vasca Navale 84, 00146, Rome, Italy}

\author{I. Gianani}
\affiliation{Dipartimento di Scienze, Universit\`a degli Studi Roma Tre, Via della Vasca Navale 84, 00146, Rome, Italy}

\author{L. Mancino}
\affiliation{Dipartimento di Scienze, Universit\`a degli Studi Roma Tre, Via della Vasca Navale 84, 00146, Rome, Italy}

\author{M. Sbroscia}
\affiliation{Dipartimento di Scienze, Universit\`a degli Studi Roma Tre, Via della Vasca Navale 84, 00146, Rome, Italy}

\author{I. Miatka}
\affiliation{Dipartimento di Matematica e Fisica, Universit\`a degli Studi Roma Tre, Via della Vasca Navale 84, 00146, Rome, Italy}

\author{F. Somma}
\affiliation{Dipartimento di Scienze, Universit\`a degli Studi Roma Tre, Via della Vasca Navale 84, 00146, Rome, Italy}

\author{M. Barbieri}
\affiliation{Dipartimento di Scienze, Universit\`a degli Studi Roma Tre, Via della Vasca Navale 84, 00146, Rome, Italy}

\begin{abstract}
Classical entanglement is a powerful tool which provides a neat numerical estimate for the study of classical correlations. Its experimental investigation, however, has been limited to special cases. Here, we 
demonstrate that the experimental quantification of the level of classical entanglement can be carried out in more general instances. Our approach enables the extension to arbitrarily shaped transverse modes and hence delivering a suitable quantification tool to describe concisely the modal structure. 
\end{abstract}

\pacs{Valid PACS appear here}% PACS, the Physics and Astronomy
\maketitle

\section{\label{sec:level1}Introduction}
The propagation of light may exhibit a strong correlation between its propagation direction or, equivalently, its transverse profile, and its polarisation, a phenomenon sometimes deemed spin-orbit coupling of light \cite{Bliokh2015}. 

This behaviour is observed in uniaxial crystals \cite{Brasselet2009}, in scanning single fluorescent microbeads with a strongly focused intensity distribution \cite{Bokor2005} and in polarisation changes along optical fibres \cite{Nakazawa1983}.
Such coupling does not allow to write the electric field separating the polarisation from the propagation; this has recently been interpreted as a manifestation of so-called classical entanglement, as it bears close resemblance to quantum two-particle system in a non-factorable state \cite{Tppel2014}.  This interpretation has led to further insights and applications, such as the formulation of an inequality criterion for the non-separability of the optical spin-orbit coupling \cite{Borges2010}, the classification of Muller matrices \cite{Simon2010}, the use of multimode waveguide transverse modes to classically simulate quantum entanglement \cite{Fu2004}, and advances in high-speed kinematic sensing \cite{BergJohansen2015}; in turn, it has inspired novel approaches to quantum technologies \cite{Parigi2015,DAmbrosio2013}. Crucially this analogy has established that the same tools used to characterise quantum entanglement are apt to describe and quantify its classical counterpart as well \cite{Leuch,Silva2016,McLaren2015}.

Experimental classical entanglement measurements have been extracted when the field transverse profile can be reconducted to a discrete basis, in particular that of Laguerre-Gauss modes, by direct inspection of the correlation between such transverse modes and the polarisation \cite{Samlan}. The flexibility of such method also meets up with the estimation of classical entanglement for vector beams \cite{Ndagano2016} which is adopted by Ndagano et al. to parameterize the vector quality factor of a vortex beam. In turn, the same classical correlations have been used to probe a quantum channel affected by atmospheric noise, based on its effect on vortex beams \cite{Ndagano2}.\\ However, classical concurrence does not occur necessarily only for these particular beam shapes. If further applications are sought, then we need a general method for measuring concurrence which does not rely on the details of the beam. Here we demonstrate that a flexible measurement strategy can be implemented by simply recurring to spatially-resolved polarisation analysis of the transverse profile. Our work demonstrates that obtaining such quantum-inspired quantifiers is not experimentally demanding, and provides concise information on the nature of mode. Through our results, the use of classical concurrence as a valid figure of merit might find applicability to novel scenarios.

\section{\label{sec:level1}Theory}
A generic (complex) electric field can be expressed as a function of its transverse position $\vec{r}$ as $E_H(\vec{r}) \hat{i}+E_V(\vec{r})\hat{j}$; the more $E_H(\vec{r})$ and $E_V(\vec{r})$ will differ, the stronger the coupling will be. We are then led to adopt a description of our field in a quantum notation as a normalized superposition state

\begin{equation}
|\Phi\rangle=\alpha|\Psi_H\rangle|H\rangle+\beta|\Psi_V\rangle|V\rangle,
\label{statoluce}
\end{equation} 

where $|H\rangle$ is the horizontal polarisation, $|V\rangle$ is the vertical polarisation and $|\Psi_H\rangle$, $|\Psi_V\rangle$ are the transverse spatial mode states associated to the horizontal and vertical polarisation respectively. Further, $|\alpha|^2$  and  $|\beta|^2$  are linked to the total intensities of the two polarisation components. 

This formalism allows us to utilise all the conceptual tools that are commonplace in quantum mechanics to assess similarities and correlations. Indeed, we can decompose the vertical spatial mode $|\Psi_V\rangle$ as a superposition of the orthogonal basis $\lbrace|\Psi_H\rangle,|\Psi_H^\dagger\rangle\rbrace$

\begin{equation}
|\Psi_V\rangle=a|\Psi_H\rangle+b|\Psi_H^\dagger\rangle,
\label{stato}
\end{equation}
where we have adopted a single-mode representation for the spatial profiles. In this description, the coefficient $a$ represents the overlap of $|\Psi_H\rangle$ and $|\Psi_V\rangle$

\begin{equation}
a=\langle \Psi_H | \Psi_V \rangle=\int d\vec{r} \Psi_H(\vec{r})\Psi_V(\vec{r})e^{i(\varphi_H(\vec{r}) - \varphi_V(\vec{r}))},
\label{integ}
\end{equation}

where $\Psi_H(\vec{r})$ and $\Psi_V(\vec{r})$ are the electric field amplitudes of spatial modes of horizontal and vertical polarization and $\varphi_H(\vec{r})$, $\varphi_V(\vec{r})$ are the spatial phases. The coefficient $b$, ensures the normalisation of the state in ~\eqref{stato}. When $|a|=1$, the state in ~\eqref{statoluce} can be factored in a spatial and a polarisation part, with no correlations; when $a=0$, the two spatial profiles are completely distinct, hence the correlation achieves the maximum permitted by the choice of $\alpha$ and $\beta$. Such information is embedded in the entanglement of the state in ~\eqref{statoluce}.

If the superposition between the two field components is not perfect - for instance, due to coupling to temporal/spectral degrees of freedom- we need to introduce a more general description, making use of a density matrix associated to the state in~\eqref{statoluce}; in the basis $\lbrace|\Psi_H\rangle|H\rangle,|\Psi_H^\dagger \rangle|H\rangle, |\Psi_H\rangle|V\rangle,|\Psi_H^\dagger \rangle|V\rangle \rbrace$ the density matrix reads:

\begin{equation}
\rho=\begin{pmatrix}
|\alpha|^2 & 0 & \epsilon \alpha \beta^* a^* & \epsilon \alpha \beta^* b^* \\
0 & 0 & 0 & 0\\
\epsilon \alpha^* \beta a & 0 & |\beta|^2 |a|^2 & |\beta|^2 a b^* \\
\epsilon \alpha^* \beta b & 0 & |\beta|^2 a^* b & |\beta|^2 |b|^2 \\
\end{pmatrix}.
\label{matricedens}
\end{equation}

The diagonal terms describes the intensities of the different contributions, while the off-diagonal terms describe their coherence. In order to account for a possible loss of coherence between the $|\Psi_H\rangle|H\rangle$ and $|\Psi_V\rangle|V\rangle$ modes, we have introduced a parameter  $0\leq \epsilon \leq 1$. The intimate analogy of this mode's structure and the quantum state of two entangled particles, suggests to evaluate classical entanglement using the quantifiers of entanglement pertinent to the density matrix $\rho$; in our $4\times4$ case, the standard choice is verifying the negativity of $\rho^{PT}$, i.e. the matrix obtained by transposing the polarisation subsystem only. The occurence of a negative eigenvalue $\lambda^{(-)}$ of $\rho^{PT}$ gives a rigorous proof of entanglement \cite{NPT}, and this can be used to define the negativity: 
\begin{equation}
N(\rho)=-2 \lambda^{(-)},
\label{negpt}
\end{equation}
which is a good entanglement quantifier, ranging from 0 for separable states, to 1 for maximal entanglement.
The negativity so-defined  is related to measurable quantities; experimentally it requires reconstructing the relative phase $\varphi(\vec{r})=\varphi_H(\vec{r}) - \varphi_V(\vec{r})$, since we keep one polarisation mode as our phase reference at any point. Contrary to shearing interferometry \cite{LSI} we do not need an absolute phase measurement; thus to obtain the relative phase $\varphi(\vec{r})$ throughout the spatial mode, a full state reconstruction is performed. An algorithm based on the Lvovsky iterative maximum-likelihood reconstruction \cite{Lvovsky} is been implemented in order to evaluate the ensemble's density matrices from statistical data. 
We used this fast reconstruction algorithm to obtain the expression of the local polarisation states; this ensures that in each point the polarisation is represented by a physically meaningful state:
\begin{equation}
\rho_{\text{pol}}=\begin{pmatrix}
\eta_H & \epsilon e^{i\varphi(\vec{r})} \sqrt{\eta_H \eta_V}\\
\epsilon^* e^{-i\varphi(\vec{r})} \sqrt{\eta_H \eta_V} & \eta_V\\
\end{pmatrix},
\label{matricedenspol}
\end{equation}
where $\eta_H$ and $\eta_V$ are the horizontal and vertical population of polarisation. 
In order to find the ensemble $\rho_{\text{pol}}$ which maximises the likelihood, we calculate the operator
\begin{equation}
R(\rho_{\text{pol}})=\sum_p \frac{I_p}{\text{pr}_p} \Pi_p,
\label{Rmatrix}
\end{equation}
where $I_p$ (with $p=\lbrace H,V,D,A,R,L \rbrace$) are the experimental intensities, 
\begin{subequations}
\begin{equation}
\Pi_H=|H\rangle\langle H|=\begin{pmatrix}
1 & 0\\
0 & 0\\
\end{pmatrix},
\Pi_V=|V\rangle\langle V|=\begin{pmatrix}
0 & 0\\
0 & 1\\
\end{pmatrix},
\end{equation}
\begin{equation}
\Pi_D=|D\rangle\langle D|=\frac{1}{2}\begin{pmatrix}
1 & 1\\
1 & 1\\
\end{pmatrix},
\Pi_A=|A\rangle\langle A|=\frac{1}{2}\begin{pmatrix}
1 & -1\\
-1 & 1\\
\end{pmatrix},
\end{equation}
\begin{equation}
\Pi_R=|R\rangle\langle R|=\frac{1}{2}\begin{pmatrix}
1 & -i\\
i & 1\\
\end{pmatrix},
\Pi_L=|L\rangle\langle L|=\frac{1}{2}\begin{pmatrix}
1 & i\\
-i & 1\\
\end{pmatrix},
\end{equation}
\end{subequations}
denote the projector operators and
\begin{equation}
\text{pr}_p=\langle p|\rho_{\text{pol}}|p\rangle=\text{Tr}(\Pi\rho_{\text{pol}})
\end{equation}
are the probabilities of the outcome polarisation state $|p\rangle$.\\
After a sufficient number of iterations to reach convergence, we attain a density matrix $\rho_{\text{pol}}$, representing the reconstructed polarisation state from which we can extract the values of the phase $\varphi(\vec{r})$ and the coherence $\epsilon$.

\section{\label{sec:level1}Experiment}
\begin{figure}[ht!]
\begin{center}
\includegraphics[width=0.45\textwidth]{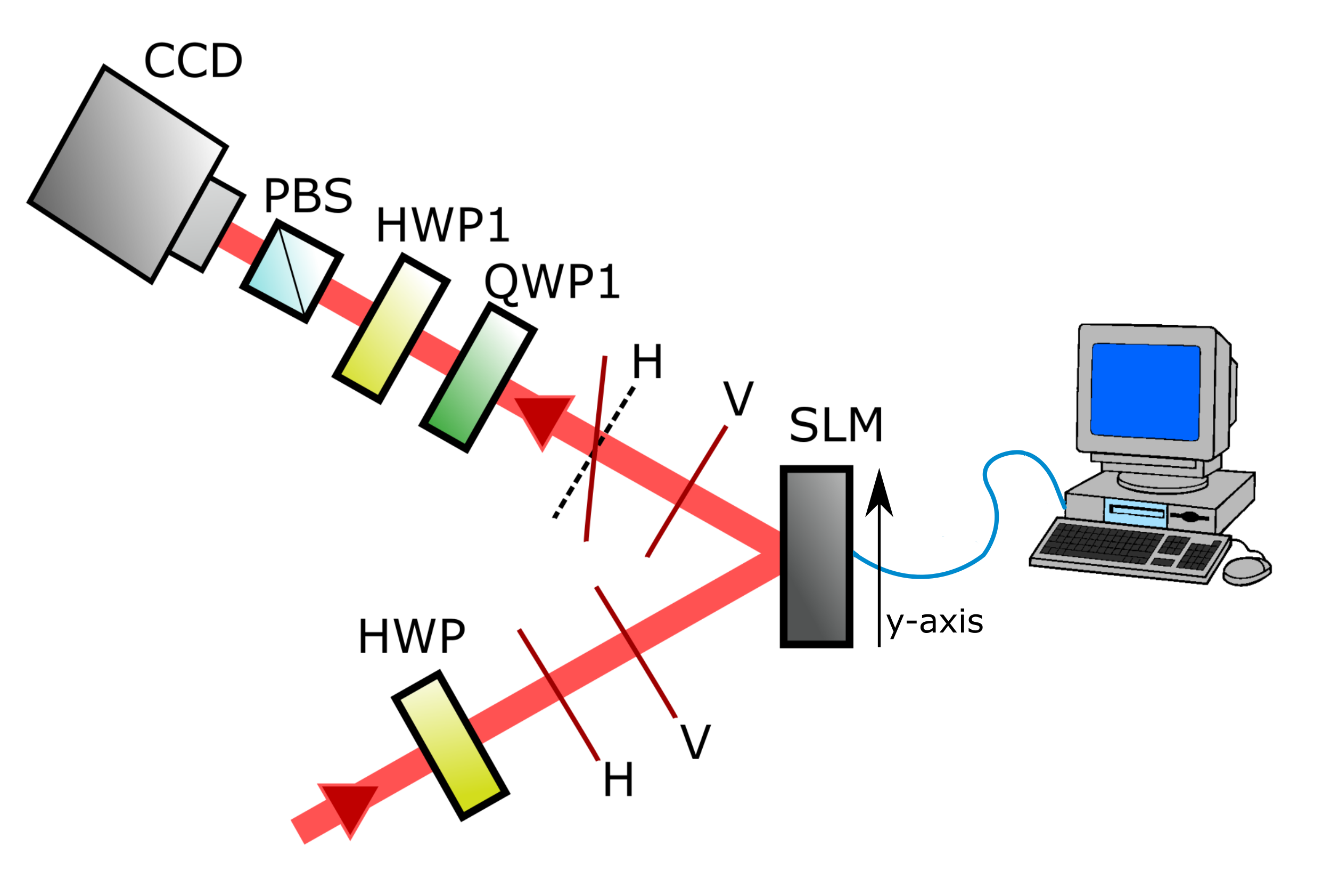}
\caption{Experimental setup. A diode laser at 810 nm, with an average power $P\sim 100$ nW prepared in the diagonal polarisation state $|D\rangle$ with a first half-wave plate (HWP) is sent on a spatial light modulator (SLM); to ensure the correct functioning of the device, the incidence angle on the SLM is set to $\theta=$ 7.1\si{\degree}. The SLM is composed by $792 \times 600$ pixels which can be independently set on a grey scale varying from 0 to 255 which maps a spatial phase onto the 0-$2\pi$ range to the wavefront with horizontal polarisation (H), preserving the vertical polarisation wavefront (V) \cite{SLM}. The phase-shaped beam is then measured in polarisation with a standard apparatus composed by a quarter-wave plate (QWP1), an half-wave plate (HWP1) and a polarising beam splitter (PBS). The beam is then detected with a CCD camera with resolution $1280 \times 1024$ pixels.}
\label{setuparticolo2}
\end{center}
\end{figure}

Our experiment consisted in measuring the negativity of shaped input fields by means of a spatially-resolved polarisation analysis, Fig. \ref{setuparticolo2}. We start taking the output of single-mode fibre, thus ensuring we initially work with a Gaussian mode. Its polarisation is prepared in the diagonal state $|D\rangle=1/\sqrt{2}(|H\rangle+|V\rangle)$, uniformly across its mode. The beam is then directed onto a spatial light modulator (SLM), which realises the coupling between the transverse position and the polarisation. Indeed, at any given point, the SLM imparts a phase $\varphi(\vec{r})$ between the horizontal and vertical polarisations in a controlled way, as demonstrated at the single-photon level in \cite{Lemos2014}. We have set the SLM in order to obtain two different configurations: a phase gradient along the $y$-axis, and a phase discontinuity in the $y$-axis.

For each SLM setting we recorded on a CCD camera six images corresponding to six different polarisations: $|H\rangle$, $|V\rangle$, $|D\rangle$, $|A\rangle=1/\sqrt{2}(|H\rangle-|V\rangle)$, $|R\rangle=1/\sqrt{2}(|H\rangle+i|V\rangle)$, $|L\rangle=1/\sqrt{2}(|H\rangle-i|V\rangle)$. Polarisation measurements are implemented through a standard apparatus consisting of an half-wave plate (HWP), a quarter-wave plate (QWP) and a polarising beam splitter (PBS). From the image corresponding to the polarisation $|p\rangle$  we extract the average intensity $I_p(j,k)$ where the couple $(j,k)$ identify a cluster of $5\times5$ camera pixels. The choice of clustering stems from the need of reducing the computational time, however the size of the clusters remains smaller than the characteristic features of the spatial modes.
Each intensity pattern is such that its integral delivers the intensity of the selected polarisation component. The linearity of the device has been tested by verifying Malus's law on a linearly polarised beam. We have adopted a cut-off for intensities lower than $1\%$ of the height of the total intensity peak as this corresponds to the noise floor.
\begin{figure}[ht!]
 \begin{center}
 \includegraphics[width=0.5\textwidth]{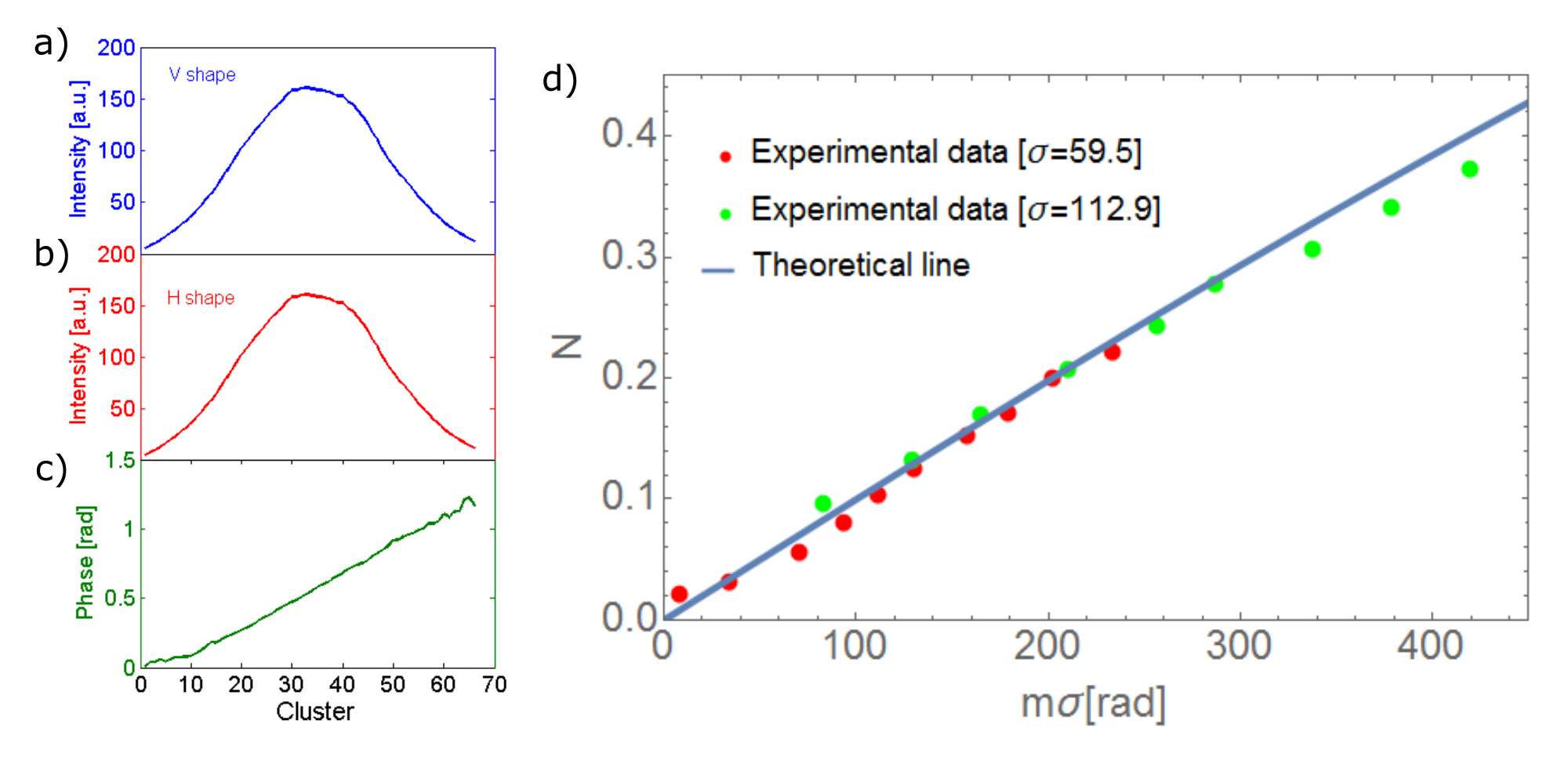}
\caption{Experimental case of polarisation-mode coupling with the phase gradient. a) Intensity shape of the V polarisation reconstructed mode. b) Intensity shape of the H polarisation reconstructed mode. c) Spatial phase gradient of the reconstructed beam mode ($m=3.913$ mrad/pixel). d) Theoretical prediction (solid line) and experimental values of the negativity for two different sizes of the beam mode (red and green dots) identified by the standard deviation $\sigma$ of the gaussian beam intensity profile. Error bars are smaller than the point size.}
 \label{grad}
 \end{center}
\end{figure}

We consider first the case of the phase gradient: Fig. \ref{grad}a-c shows a typical mode reconstruction where we plot a slice of the intensities and phase for the sake of clarity. The spatial shape of the beam is unaffected by the action of the SLM for both polarisation; the sheer result is the introduction of a position-dependent phase whose behaviour exhibits the expected linear trend. We have applied gradients with different slopes $m$ to our beam and evaluated the negativity for every setting. The dispersion of the measured coherence values $\epsilon(j,k)$ is beneath $3\%$: our description~\eqref{matricedens} is then well justified. The mean value $\epsilon$ in $\rho$ is then taken as the average value weighted with the measured intensities \cite{epsilon}. The results, shown in Fig. \ref{grad}d together with the theoretical prediction, lie close to the ideal case \cite{note}. Discrepancies can be attributed to unexpected features of the beam, likely originated by the other optical elements.

\begin{figure}[ht!]
 \begin{center}
 \includegraphics[width=0.5\textwidth]{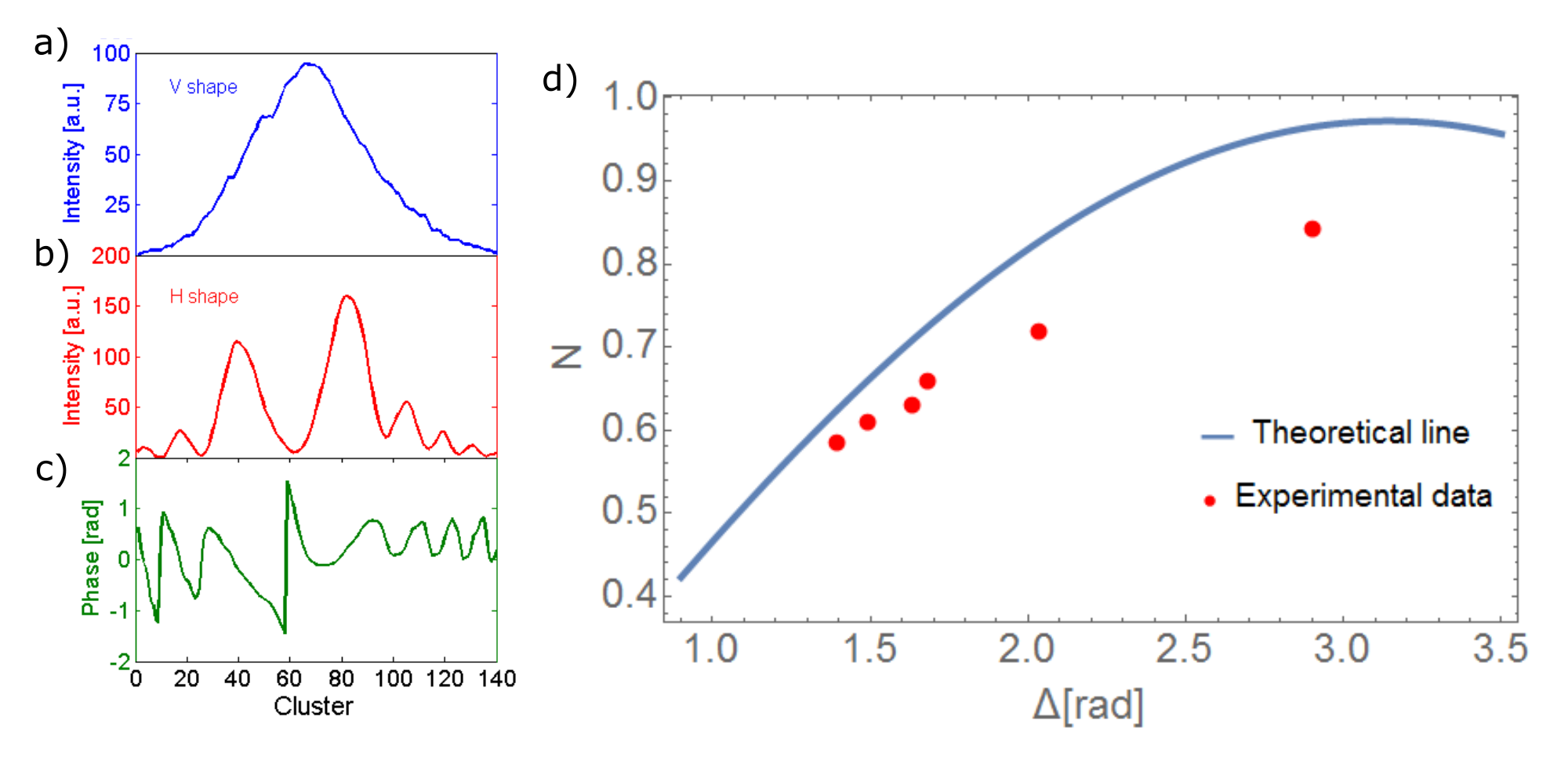}
\caption{Experimental case of polarisation-mode coupling with a phase jump. a) Intensity shape of the V polarisation reconstructed mode. b) Intensity shape of the H polarisation reconstructed mode. c) Spatial phase jump of the reconstructed beam mode ($\Delta=1.394$ rad). d) Theoretical prediction (solid line) and experimental values (red dots) of the negativity. Error bars are smaller than the point size.}
 \label{jump}
 \end{center}
\end{figure}

We apply our method to a more structured mode in which the image of the SLM presents a phase discontinuity $\Delta$. The reconstruction of a typical case is reported in Fig. \ref{jump}a-c where it can be seen that, while the V mode is left untouched, the H mode suffers a diffraction from the sharp edge. This is also reflected in the rich phase structure that we have reconstructed. In Fig. \ref{jump}d we plot the reconstructed negativities for different phase jumps; the overall trend resembles that of the simple prediction (solid line) that only considers the phase jump but not the intensity and phase modulations. By our method we can quantify the impact of such features on the coupling of mode and polarisation. This is illustrative of the effectiveness of condensing the information on such coupling into one parameter.

\section{\label{sec:level1}Conclusions}
Concluding, we have demonstrated that finding a quantitative evaluation of the mode-polarisation coupling through classical entanglement is feasible in general instances performing a complete polarisation tomography typical of the quantum approach. This allows, by isolating a single facet of the problem, to obtain a meaningful numeric figure as a result. We believe that this approach could be helpful in a diverse range of applications where the complexity of the problem is higher. This may be the case especially when dealing with polarisation-transforming processes which depend on the spatial position \cite{Pierangelo}. We think that the experimental simplicity, conciseness, and informativeness of approaching classical coherence with quantum-inspired methods will establish it as the standard analysis technique. 

\section{\label{sec:level1}Acknowledgements}
The authors would like to thank Prof. Fabio Sciarrino for the loan of scientific equipment. This work has been partially supported by the European Commission, under the programme Horizon 2020, grant number 665148. Marco Barbieri has been supported by a Rita Levi-Montalcini fellowship of MIUR.

% Bibliography

%\bibliographystyle{prarev4-1} % Tell bibtex which bibliography style to use
%\bibliography{bibliography}

\begin{thebibliography}{99}
\bibitem{Bliokh2015}K. Y. Bliokh, F. J. Rodr{\'{\i}}guez-Fortu{\~{n}}o, F. Nori and A. V. Zayats, Nature Phot.{\bf 9}, 796--808 (2015).
\bibitem{Brasselet2009} E. Brasselet, Y. Izdebskaya, V. Shvedov, A. S. Desyatnikov, W. Krolikowski and Y. S. Kivshar, \ol {\bf 34}, 546 (1983).
\bibitem{Bokor2005} N. Bokor, Y. Iketaki, T. Watanabe and M. Fujii, Opt. Express {\bf 13}, 10440 (2005).
\bibitem{Nakazawa1983} M. Nakazawa, M. Tokuda and Y. Negishi, \ol {\bf 8}, 546 (1983).
\bibitem{Tppel2014}F. T\"{o}ppel, A. Aiello, C. Marquardt, E. Giacobino and G. Leuchs, New J. Phys.{\bf 16}, 073019 (2014).
\bibitem{Borges2010} C. V. S. Borges, M. Hor-Meyll, J. A. O. Huguenin and A. Z. Khoury, \pra {\bf 84}, (2010).
\bibitem{Simon2010} B. N. Simon, S. Simon, F. Gori, M. Santarsiero, R. Borghi, N. Mukunda and R. Simon, \prl {\bf 104}, (2010).
\bibitem{Fu2004} J. Fu, Z. Si, S. Tang and J. Deng, \pra {\bf 70}, (2004).
\bibitem{BergJohansen2015} S. Berg-Johansen {\it et al.}, Optica {\bf 2}, 864 (2015).
\bibitem{Parigi2015} V. Parigi, V. D'Ambrosio, C. Arnold, L. Marrucci, F. Sciarrino and J. Laurat, Nature Communications {\bf 6}, 7706 (2015).
\bibitem{DAmbrosio2013} V. D'Ambrosio {\it et al.}, Nature Communications {\bf 4}, (2013).
\bibitem{Leuch} A. Aiello, F. T{\"o}ppel, C. Marquardt, E. Giacobino and G. Leuchs, {arXiv:1409.0213}, (2014).
\bibitem{McLaren2015} M. McLaren, T. Konrad and A. Forbes, \pra {\bf 92}, 023833 (2015).
\bibitem{Silva2016} B. Pinheiro da Silva, M. Astigarreta Leal, C. E. R. Souza, E. F. Galv{\~{a}}o, A. Z. Khoury, J. Phys. B: At. Mol. Opt. Phys. {\bf 49}, 055501 (2016).
\bibitem{Samlan} C. T. Samlan and N. K. Viswanathan, {arXiv:1506.07112}, (2015).
\bibitem{Ndagano2016} B. Ndagano, H. Sroor, M. McLaren, C. Rosales-Guzm{\'{a}}n and A. Forbes, \ol~{\bf 41}, 3407 (2016).
\bibitem{Ndagano2} Bienvenu Ndagano {\it et al.}, {arXiv:1605.05144}, (2016).
\bibitem{NPT} J. Lee, M. S. Kim, Y. J. Park and S. Lee, J. Mod. Opt.{\bf 47}, 2151--2164 (2000).
\bibitem{LSI} S. Velghe, J. Primot, N. Gu{\'{e}}rineau, M. Cohen and B. Wattellier, \ol{\bf 30}, 245 (2005).
\bibitem{Lvovsky} A. I. Lvovsky, J. Opt. B: quantum and semiclassical optics, S556-S559 (2004).
\bibitem{SLM} Hamamatsu, \texttt{https://www.hamamatsu.com/resources/pdf/ssd /e12\char`_ handbook\char`_ lcos\char`_ slm.pdf}.
\bibitem{Lemos2014} G. Barreto Lemos, J. O. de Almeida, S. P. Walborn, P. H. Souto Ribeiro and M. Hor-Meyll, \pra {\bf 89}, (2014).
\bibitem{epsilon}Typical values of $\epsilon$ range from $0.97$ to $0.99$ for both experimental cases.
\bibitem{note}The error on the negativity has been evaluated through a Monte Carlo routine: the analysis is repeated by changing the values of the observed intensities in each pixel within $1\%$ of the measured maximum value. This is consistent with the observed statistics on a repeated acquisition of one image.
\bibitem{Pierangelo} M. R. Antonelli {\it et al.}, Opt. Express {\bf 18}, 10200--10208 (2010).
\end{thebibliography}
\end{document}